\documentclass[aps,pra,eqsecnum,superscriptaddress,showpacs,floatfix,nofootinbib]{revtex4-1}
\usepackage{graphicx}
\usepackage{bm}
\usepackage{dsfont}
\usepackage{amsmath,amsfonts,amssymb}
\usepackage{float}
\usepackage{hyperref}
\usepackage{amsmath}
\usepackage{amsfonts}
\usepackage{amssymb}

\hypersetup{colorlinks=true,linkcolor=blue,citecolor=blue,urlcolor=blue}

\begin{document}

\title{Probing Sound Speed of an Optically-Trapped Bose Gas with Periodically Modulated Interactions by Bragg Spectroscopy}

\author{Lei Chen}
\affiliation{Shenyang National Laboratory for Materials Science,
Institute of Metal Research, Chinese Academy of Sciences, Wenhua
Road 72, Shenyang 110016, China}

\author{Wu Li}
\affiliation{Shenyang National Laboratory for Materials Science,
Institute of Metal Research, Chinese Academy of Sciences, Wenhua
Road 72, Shenyang 110016, China}

\author{Zhu Chen}
\affiliation{National Key Laboratory of Science and Technology on Computational Physics, Institute
of Applied Physics and Computational Mathematics, Beijing 100088, China}

\author{Zhidong Zhang}
\affiliation{Shenyang National Laboratory for Materials Science,
Institute of Metal Research, Chinese Academy of Sciences, Wenhua
Road 72, Shenyang 110016, China}

\author{Zhaoxin Liang}
\email{Corresponding author: zhxliang@imr.ac.cn}
\affiliation{Shenyang National Laboratory for Materials Science,
Institute of Metal Research, Chinese Academy of Sciences, Wenhua
Road 72, Shenyang 110016, China}

\begin{abstract}
A Bose-Einstein condensate (BEC) with periodically modulated interactions
(PMI) has emerged as a novel kind of periodic superfluid, which has been
recently experimentally created using optical Feshbach resonance. In this
paper, we are motivated to investigate the superfluidity of a BEC with PMI
trapped in an optical lattice (OL). In particular, we explore the effects of
PMI on the sound speed and the dynamical structure factor of the model
system. Our numerical results, combined with the analytical results in both
the weak-potential limit and the tight-binding limit, have shown that the
PMI can strongly modify the sound speed of a BEC. Moreover, we have shown
that the effects of PMI on sound speed can be experimentally probed via the
dynamic structure factor, where the excitation strength toward the first
Bogoliubov band exhibits marked difference from the non-PMI one. Our
predictions of the effects of PMI on the sound speed can be tested using the
Bragg spectroscopy.
\end{abstract}

\maketitle

\section{Introduction}

By using the optical Feshbach resonance (OFR), a Bose-Einstein condensate
(BEC) with periodically modulated interactions (PMI) has been recently
realized in the experiments \cite{1,2,3,4}. Such a novel periodic
superfluid, which has no analogue in condensed matter physics, has opened up
new avenues to exploring the superfluidity of quantum many-body systems with PMI.

Meanwhile, there exists another conventional way of creating a periodic
superfluid via loading a BEC into an optical lattice (OL) \cite{5,6,7,8,9,10,11,12,13,14}. Both being periodic, however, a BEC with PMI and an optically trapped BEC
have exhibited interestingly different superfluid behavior. For example,
it's well known that the dynamic instability plays a key role in destroying
the superfluidity in a periodic superfluid \cite{15,16,17,18,19,20,21,22,23}. In this context, Ref. \cite{24} has
found that, in a BEC with PMI, all Bloch waves in the lowest band will
inevitably become dynamical unstable when the PMI is strong enough; whereas
in comparison, an optically trapped BEC in the lowest band will be more
stable with increasing interaction. Inspired by such comparisons, we are
interested in the case when a BEC is in the presence of both PMI and an
optical lattice. Here we investigate the effect of PMI on the sound speed of
an optically trapped BEC, and discuss its exploration via measuring the
dynamic structure factor using the Bragg spectroscopy \cite{25,26,27,28,29,30,31}. The motivation is twofold. First, the
sound speed is intimately related to the concept of superfluidity and its
exploration. Second, the application of Bragg spectroscopy in such a novel
kind of periodic superfluid is itself worthy of more efforts.

The main purpose of this work is to theoretically investigate both the sound
speed and the dynamical structure factor of a BEC with PMI trapped in an OL
\cite{32,33} using the mean-field theory. Our results
show that, compared to the non-PMI counterpart, (i) the PMI can
significantly affect the sound speed; (ii) the excitation strength toward
the first Bogoliubov band in a BEC with PMI is markedly difference from the
non-PMI one. Based on these calculations, we also discuss the conditions for
possible experimental realizations of our scenarios.

The paper is organized as follows. First, in Sec.~\ref{sec:Model} we derive
the effective model for a quasi-one-dimensional BEC with PMI in an OL. Then
in Sec.~\ref{sec:Methods}, we study the sound propagation and dynamic
structure factor of the model system in different parameter regimes, using
both analytical and numerical approaches. Finally, we summarize our results
in Sec.~\ref{sec:Summary} and present an outlook.

\section{The Model System}

\label{sec:Model}

\begin{figure}[tbp]
\begin{centering}
\includegraphics[scale=0.4]{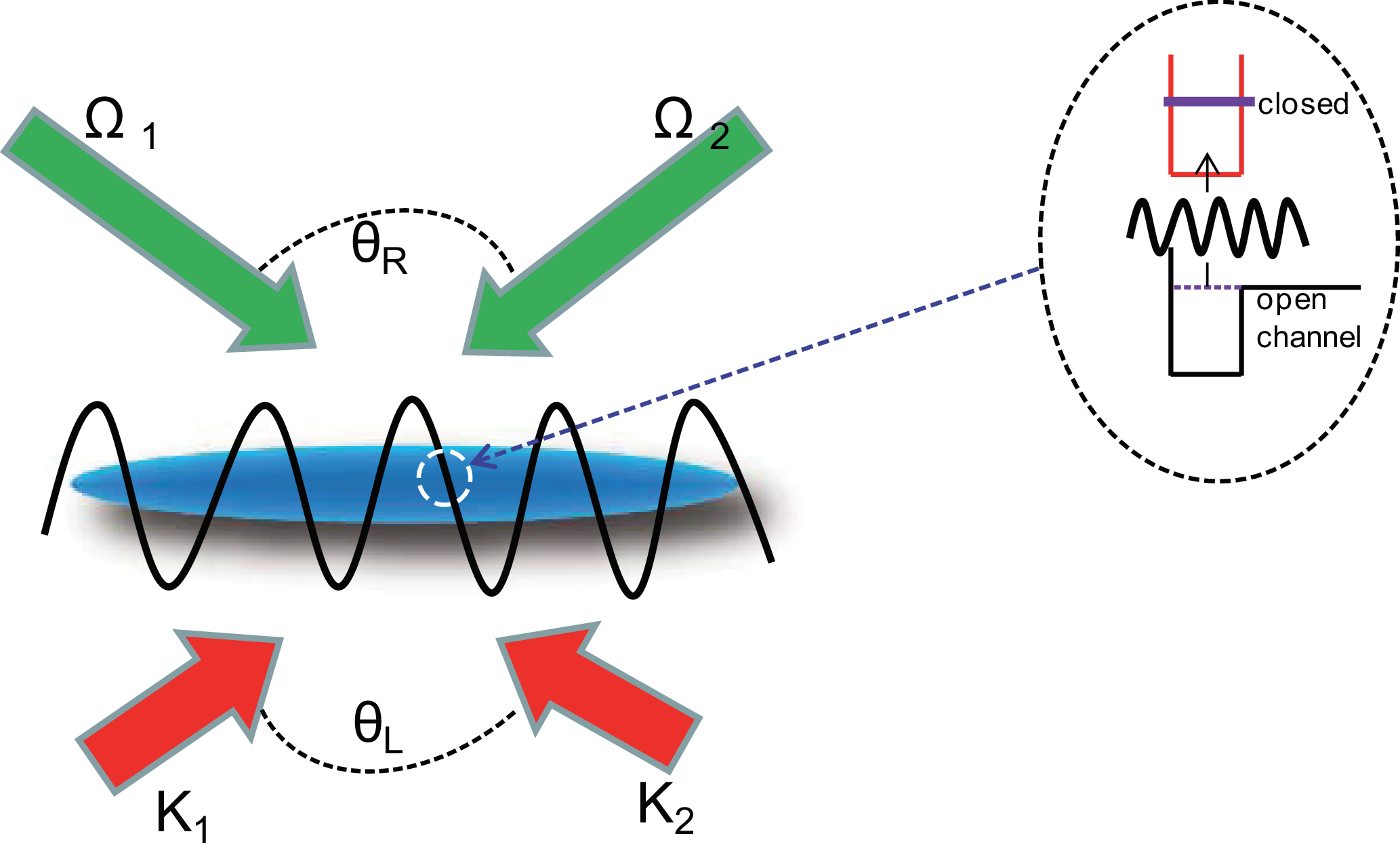}
\par\end{centering}
\caption{(Color online) Schematic setup of a BEC with PMI in an OL. Two
laser beams of $\Omega _{1}$ and $\Omega _{1}$ generate the periodically
modulated interaction by OFR; while the other two beams of $k_{1}$ and $k_{2}
$ generate the OL. }
\label{model}
\end{figure}

We consider a BEC with PMI by OFR trapped in a strongly anisotropic lattice
potential as shown in Fig. \ref{model}. Specifically, the transverse lattice
confinement is tuned sufficiently strong to freeze the atomic motion in
these directions such that atoms are only allowed to tunnel in the $x$-direction,
leading to the realization of a quasi-one-dimensional geometry \cite{34,35}. The OL
along the $x$-direction reads $V\cos (2k_{L}x)$ with $V$ being the lattice
strength. The wave vector of the lattice $k_{L}=2\pi \sin (\theta
_{L}/2)/\lambda _{L}$ can be manipulated via the wavelength of the lasers $%
\lambda _{L}$ and the angle $\theta _{L}$ between the two lasers. The PMI in
the form of $g_{1}+g_{2}\cos \left( 2k_{Z}x\right) $ for a BEC has been
experimentally realized using OFR. Here, $g_{1}$ and $g_{2}$ are positive
parameters and $k_{Z}=2\pi \sin \left( \theta _{R}/2\right) /\lambda _{Z}$
with $\lambda _{Z}$ being the wavelength of the OFR and $\theta _{R}$ being
the angle between the OFR beams (see Fig. \ref{model}). Note that $g_{1}$, $%
g_{2}$ and $k_{Z}$ can all be tuned experimentally by adjusting OFR laser
beams. At the mean-field level, our model system can be well described by
the Gross-Pitaevskii (GP) Eq. \cite{24},
\begin{equation}
i\hbar \frac{\partial \psi }{\partial t}=-\frac{\hbar ^{2}}{2m}\frac{%
\partial ^{2}}{\partial x^{2}}\psi +V\cos \left( 2k_{L}x\right) \psi +\left[
g_{1}n_{0}+g_{2}n_{0}\cos \left( 2k_{Z}x\right) \right] \left\vert \psi
\right\vert ^{2}\psi ,  \label{TGP}
\end{equation}%
where $m$ is the atom mass, $\psi$ is the condensate wave function and $n_{0}$
is the average condensate density. While both $k_{L}$ and $k_{Z}$ can be
tuned, as a first step to investigating the superfluidity of the novel
periodic superfluid under consideration, we will limit ourselves to the case
$k_{L}=k_{z}$ throughout this paper. The corresponding 1D GP Eq. (\ref{TGP})
reads in a dimensionless form as
\begin{equation}
i\hbar \frac{\partial \psi }{\partial t}=-\frac{1}{2}\frac{\partial ^{2}}{%
\partial x^{2}}\psi \left( x\right) +v\cos \left( x\right) \psi \left(
x\right) +\left[ c_{1}+c_{2}\cos \left( x\right) \right] \left\vert \psi
\left( x\right) \right\vert ^{2}\psi \left( x\right) .  \label{eq:2}
\end{equation}%
In Eq. (\ref{eq:2}), the energy unit is $8E_{R}$ with $E_{R}=2\hbar ^{2}\pi
^{2}/m\lambda _{L}^{2}$ being the recoil energy and the length unit is $%
1/2k_{L}$. The lattice strength and nonlinear coefficients scale as: $%
v=V/\left( 8E_{R}\right) $, $c_{1}=g_{1}n_{0}/\left( 8E_{R}\right) $ and $%
c_{2}=g_{2}n_{0}/\left( 8E_{R}\right) $.

In this work, we are interested in (i) the sound speed in the novel periodic
superfluid described by Eq. (\ref{eq:2}); (ii) the probe of sound speed by
using the Bragg spectroscopy, where the dynamic structure factor of the model
system is directly measured. Before we proceed into concrete calculations,
let us first present a general framework concerning the sound speed and the
dynamic structure factor of the system under consideration:

(i) It has been known that the sound propagation and its speed of an
optically trapped BEC can be discussed from two perspectives \cite{36}.
In one, the sound speed is viewed as a quantity intimately related to the
superfluidity of a BEC and its macroscopic dynamics, the definition \cite{36,37,38}
\begin{equation}
c_{s}=\sqrt{\frac{1}{\kappa m^{\ast }}},  \label{Soundspeed}
\end{equation}%
with the compressibility $\kappa$ and the effective mass $m^{\ast }$ being
defined as follows,
\begin{equation}
\frac{1}{m^{\ast }}=\lim_{k\rightarrow 0}\frac{d^{2}\epsilon _{k}}{dk^{2}}%
,\qquad \kappa ^{-1}=n_{0}\frac{\partial \mu }{\partial n_{0}}.
\label{mstar}
\end{equation}%
Here, the chemical potential $\mu $ reads $\mu =\partial (n_{0}\epsilon
_{k})/\partial n_{0}$ with the energy per particle $\epsilon _{k}$ being
written as,
\begin{equation}
\epsilon _{k}=\frac{1}{2\pi }\int_{-\pi }^{\pi }dx\left[ \frac{1}{2}%
\left\vert \frac{\partial \psi }{\partial x}\right\vert ^{2}+v\cos (x)|\psi
|^{2}+\frac{1}{2}\left( c_{1}+c_{2}\cos (x)\right) \left\vert \psi
\right\vert ^{4}\right] .  \label{Ek}
\end{equation}

On the other hand, the sound propagation in a BEC can be also treated as a
long wavelength response to an external perturbation, and the corresponding
speed can be calculated using Bogoliubov theory. In more details, the
low-energy excitation in connection to the sound propagation can be
described by a small perturbation to the wave function in Eq. (\ref{eq:2})
as $\psi (x,t)=[\psi _{0}(x)+\delta \psi (x,t)]\exp (-i\mu t)$, with $\psi
_{0}$ being the ground state. By decomposing $\delta \psi (x,t)=u(x)\exp
(iqx-i\omega t)+v^{\ast }(x)\exp (-iqx+i\omega t)$, together with Eq. (\ref%
{eq:2}), we obtain Bogoliubov-de Gennes (BdG) equations reading
\begin{eqnarray}
&&\left[ -\frac{1}{2}\frac{\partial ^{2}}{\partial x^{2}}+v\cos (x)-\mu
+2\left( c_{1}+c_{2}\cos (x)\right) |\psi |^{2}\right] u_{jq}+\left(
c_{1}+c_{2}\cos (x)\psi ^{2}\right) v_{jq}=\omega _{j}(q)u_{jq}  \notag \\
&&\left[ -\frac{1}{2}\frac{\partial ^{2}}{\partial x^{2}}+v\cos (x)-\mu
+2\left( c_{1}+c_{2}\cos (x)\right) |\psi |^{2}\right] v_{jq}+\left(
c_{1}+c_{2}\cos (x)\psi ^{\ast 2}\right) u_{jq}=-\omega _{j}(q)v_{jq}.
\label{BE}
\end{eqnarray}%
Here, $u$ and $v$ are the Bogoliubov amplitudes that satisfy the
normalization and orthogonality conditions $\int \left[ u_{j^{^{\prime
}q}}^{\ast }(x)u_{jq}(x)-v_{j^{^{\prime }q}}^{\ast }(x)v_{jq}(x)\right]
dx=\delta _{j^{^{\prime }},j}$; the $q$ and $\omega _{j}\left( q\right) $
are the wave vector and the energy of the Bogoliubov excitations with $j$
being the band index, respectively. According to Eq. (\ref{BE}), the sound
speed of a quasi-1D BEC under consideration can then be defined as
\begin{equation}
c_{s}=\lim_{{q}\rightarrow 0}\frac{\omega _{j=1}\left( q\right) }{q}.
\label{Soundspeed0}
\end{equation}%
These two definitions on the sound speed (Eqs. (\ref{Soundspeed0}), (\ref%
{Soundspeed})) have been proved equivalent \cite{36} for an optically
trapped BEC.

(ii) In the second scenario, the Bragg spectroscopy measures the energy
spectrum of the model system by stimulating small-angle light scattering,
delivering a momentum $p$ and an energy $\omega $. Theoretically, such light
scattering directly corresponds to the dynamic structure factor of $%
S(p,\omega )$, which is the Fourier transform of density-density
correlations function of the system. For a periodic superfluid, there exist
an infinite set of $\omega _{j}$ for each $q$ in the first Brillouin zone,
forming a Bogoliubov band labeled by index $j$. Hence, when the external
probe $p$ is varied, an infinite set of excitation strength $Z_{j}(p)$
relative to the $j$-th Bogoliubov band are excited, reading,

\begin{equation}
Z_{j}(p)=\left\vert \int \left[ u_{jq}^{\ast }+v_{jq}^{\ast }\right]
e^{ipx}\psi _{0}(x)dx\right\vert ^{2}, \label{Zj}
\end{equation}

where $q$ lies in the first Brillouin zone and is fixed by the relation $%
q=p+l2\pi /d$. Summing up all the $Z_{j}(p)$ in Eq. (\ref{Zj}) under the
condition of energy conservation, we obtain the dynamic structure factor as
follows,
\begin{equation}
S(p,\omega )=\sum_{j}Z_{j}(p)\delta (\omega -\omega _{j}(p)),  \label{DSF}
\end{equation}%
The sound speed of a BEC can be extracted directly from the slope of the
linear part of the excitation spectrum by employing the Bragg spectroscopy.
In more details, the Bragg spectroscopy in an optically-trapped BEC is
performed by superimposing a periodic traveling wave potential on the
lattice. Then, the technique of Bragg spectroscopy to measure the energy
spectrum essentially boils down to probing the dynamic structure factor,
namely the response of a BEC to an external density perturbation \cite{25,26,27,28,29,30,31,41}.

Having laid out the basic theoretical framework, below we proceed to
illustrate with an experimentally relevant system for concrete
investigations. At mean field level, Eq. (\ref{eq:2}) describes a BEC with
PMI in an OL, where the main physics is determined by three parameters, $v$,
$c_{1}$ and $c_{2}$ (respectively characterize the lattice strength, the
bare interaction strength, and the periodic interaction strength). All these
parameters can be experimentally controlled using the state-of-art
technologies. In typical experiments, the lattice strength $V$ can be turned
from $0$ to $32E_{R}$ almost at will, corresponding to the regime of $0\leq
v\leq 4$. Furthermore, both the $c_{1}$ and $c_{2}$ can be controlled in a
very versatile manner via the technology of OFR.

\section{Probing sound speed by dynamic structure factor}

\label{sec:Methods}

\subsection{Sound speed}

\begin{figure}[tbp]
\begin{centering}
\includegraphics[scale=0.8]{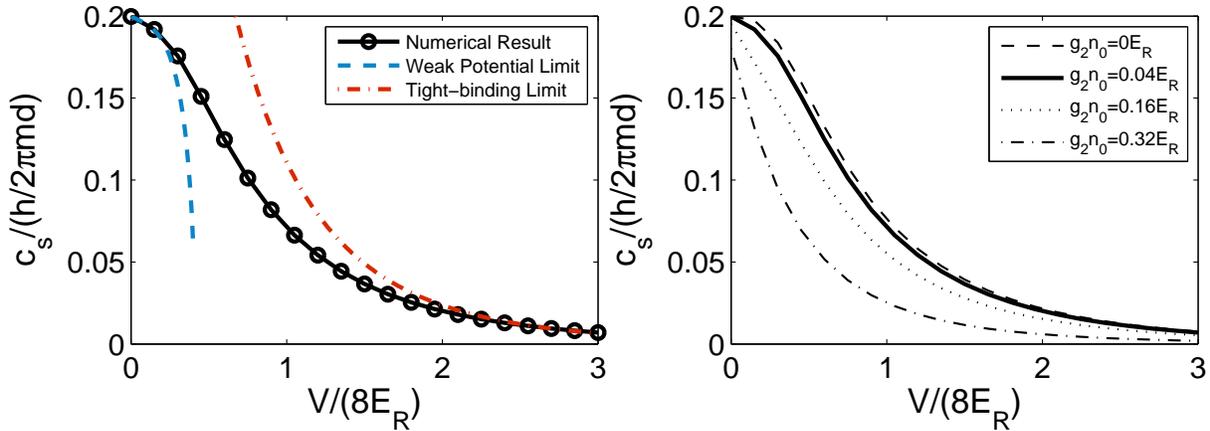}
\par\end{centering}
\caption{(Color online) Left panel: Sound speed of a BEC with PMI trapped in
an OL as a function of the lattice depth $V$ with $g_{1}n_{0}=0.32E_{R}$ and
$g_{2}n_{0}=0.04E_R$. The solid, dashed and dot-dashed lines correspond to
the numerical results, the analytical results in weak-potential limit and
tight-binding limit respectively. Right panel: The effects of $g_2n_0$ on
the sound speed via lattice strength $V$ with fixed $g_1n_0=0.32E_R$.}
\label{Sound}
\end{figure}

The previous section has set the stage for our study on the sound in a BEC
with PMI trapped in an OL. In this section, we will systematically study the
effects of PMI on the sound speed, using both analytical and numerical
methods. As we have discussed in Eqs. (\ref{Soundspeed}) and (\ref%
{Soundspeed0}), there exist two routes to calculating the sound speed. Note
that Both approaches consists in numerically solving the GP Eq. (\ref%
{eq:2}) based on the Bloch theorem. Once the Bloch waves of $\psi $ have been
found, we can (i) proceed to numerically solve BdG Eq. (\ref{BE}) and then
obtain the numerical results of sound speed according to Eq. (\ref%
{Soundspeed}); or we can (ii) calculate the energy $E(k)$ from Eq. (\ref{Ek}%
), derive the compressibility $\kappa $ and the effective mass $m^{\ast }$,
which will give the sound velocity $c_{s}$ following from Eq. (\ref%
{Soundspeed0}). In this work, we shall use both methods to calculate the
sound speed, which will be shown to agree with each other as expected.
Moreover, in order to comprehensively reveal the effects of PMI on the sound
speed, we will proceed our analysis in two steps:

(i) In the first step, we fix the PMI of $g_{2}n_{0}$, so as to figure out
how the sound speed $c_{s}$ responds to the variation of $V$. As shown in
Fig. \ref{Sound}, with the increase of $V$, the sound speed $c_{s}$
decreases monotonically, implying that the increasing effective mass $%
m^{\ast }$ always wins the competition against the decreasing
compressibility $\kappa $ in determining $c_{s}$. In order to achieve a
clearer understanding of Fig. \ref{Sound}, we have obtained analytical
results in both the weak-potential limit and tight-binding limit. In the
weak-potential limit, the analytical expression of sound speed are given by
(see Eq. (\ref{eq:11}) in Appendix)
\begin{equation}
c_{s}=\sqrt{c_{1}}\left[ 1-\frac{c_{2}^{2}}{c_{1}\left( 1+4c_{1}\right) }-%
\frac{c_{2}^{2}-v^{2}}{c_{1}\left( 1+4c_{1}\right) ^{2}}-\frac{\left(
c_{2}+v\right) ^{2}}{c_{1}\left( 1+4c_{1}\right) ^{3}}-\frac{4\left(
c_{2}+v\right) ^{2}}{\left( 1+4c_{1}\right) ^{2}}\right] .  \label{soundweak}
\end{equation}%
It's clear that if $c_{2}=0$, Eq. (\ref{soundweak}) recovers the sound
speed in an OL \cite{36,39,40} which decreases monotonically
with $v$ as it should be. In the presence of PMI with $c_{2}\neq 0$, the
second term and the last two terms in the square brackets in Eq. (\ref{eq:11}%
) are definitely negative, whereas the third term can be either positive or
negative depending on the $c_{2}$ relative to $v$. This suggests that both $%
c_{2}$ and $v$ play the same role in determining the $c_{s}$. Moreover, as
is shown by the dashed curve in Fig. \ref{Sound}, our numerical results
agree well with Eq. (\ref{soundweak}). In the opposite tight-binding limit, the
analytically derived sound speed (see Eq. (\ref{eq:15a}) in the Appendix)
also agrees well with the numerical result.

(ii) In the second step, we fix $g_{1}n_{0}$ and scan the sound speed $c_{s}$
as a function of $V$ for different choices of $g_{2}n_{0}$. As shown in the
right panel in Fig. \ref{Sound}, the sound speed behaves very differently
from the non-PMI one. Our results imply that the effect of PMI on the sound
speed can be measured within the current experimental capabilities. In what
follows, we will develop a scheme of probing the effects of PMI on the sound
speed via the dynamic structure factor (DSF).

\subsection{Dynamic Structure factor}

In order to characterize and investigate the capability of measuring the
sound speed by using the Bragg spectroscopy, we numerically calculate the
DSF of the system defined in Eq. (\ref{DSF}).

Before any further concrete calculations, we use the sum rule approach to
analyze the basic properties of the dynamic structure factor. The first sum
rule gives the static structure factor $S(p)$ by integrating the $S(p,\omega
)$ \cite{37,38}
\begin{equation}
S(p)=\int S(p,\omega )d\omega =\sum_{n}Z_{n}(p,\omega ).
\end{equation}%
We expect that $S(p)$ will be strongly affected by the combined presence of
OL and PMI. The second f-sum rule on $S(p,\omega )$ is a direct consequence
of particle conservation of the model system and represents a statement of
the conservation law, reading \cite{37,38},
\begin{equation}
\int \hbar \omega S(p,\omega )d\omega =\frac{p^{2}}{2m}.
\end{equation}%
In the long wavelength limit, the static response function manifests itself
as the response of the system density to a static force, which is
intimately related to the compressibility of the system, giving the third
sum rule \cite{37,38},
\begin{equation}
\lim_{p\rightarrow 0}\int \frac{S(p,\omega )}{\hbar \omega }d\omega =\frac{%
\kappa }{2}. \label{lim}
\end{equation}%
We emphasize that our following numerical results have been double-checked
by checking whether they satisfy the above three sum rules.

Now, we are equipped to study the effects of PMI on the sound speed by
calculating the $S(p,\omega )$. To this end, we shall focus on two
scenarios: first, we set $g_{2}n_{0}=0$ and calculate the dynamic structure
factor, which will then serve as the reference for later comparisons. Then,
we turn on the PMI ($g_{2}n_{0}\neq 0$) and study the effect of the combined
presence of PMI and optical lattice on the dynamic and static structure
factor.

\begin{figure}[tbp]
\begin{centering}
\includegraphics[scale=0.6]{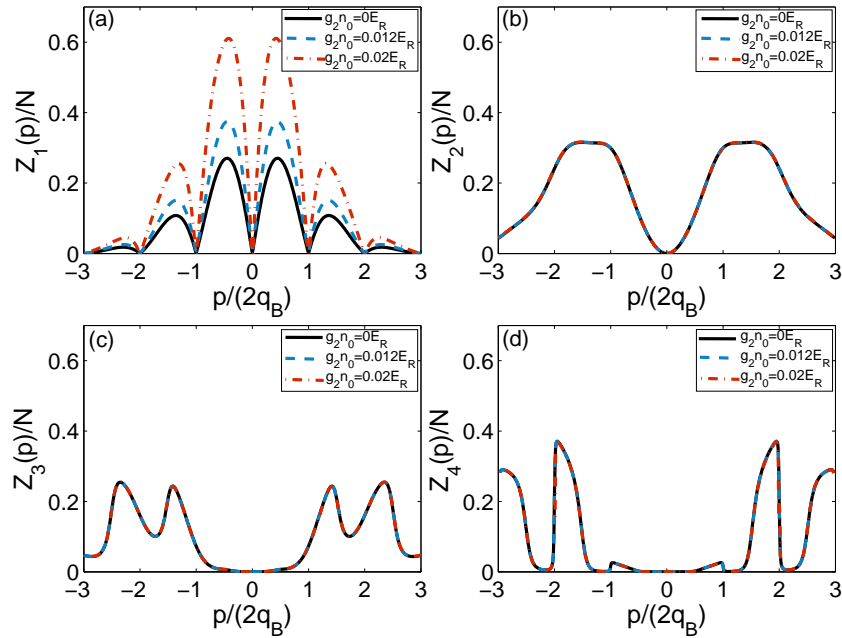}
\par\end{centering}
\caption{(Color online) The excitation strengths $Z_{j}$ $%
\left(j=1,2,3,4\right)$, as of function of momenta of an extra probe. The
full, the dashed and the dot-dashed lines corresponds to $g_{2}n_{0}=0E_{R}$%
, $g_{2}n_{0}=0.012E_{R}$ and $g_{2}n_{0}=0.02E_{R}$ with $V=10E_{R}$ and $%
g_{1}n_{0}=0.02E_{R}$, respectively.}
\label{DSF10}
\end{figure}

\begin{figure}[tbp]
\begin{centering}
\includegraphics[scale=0.6]{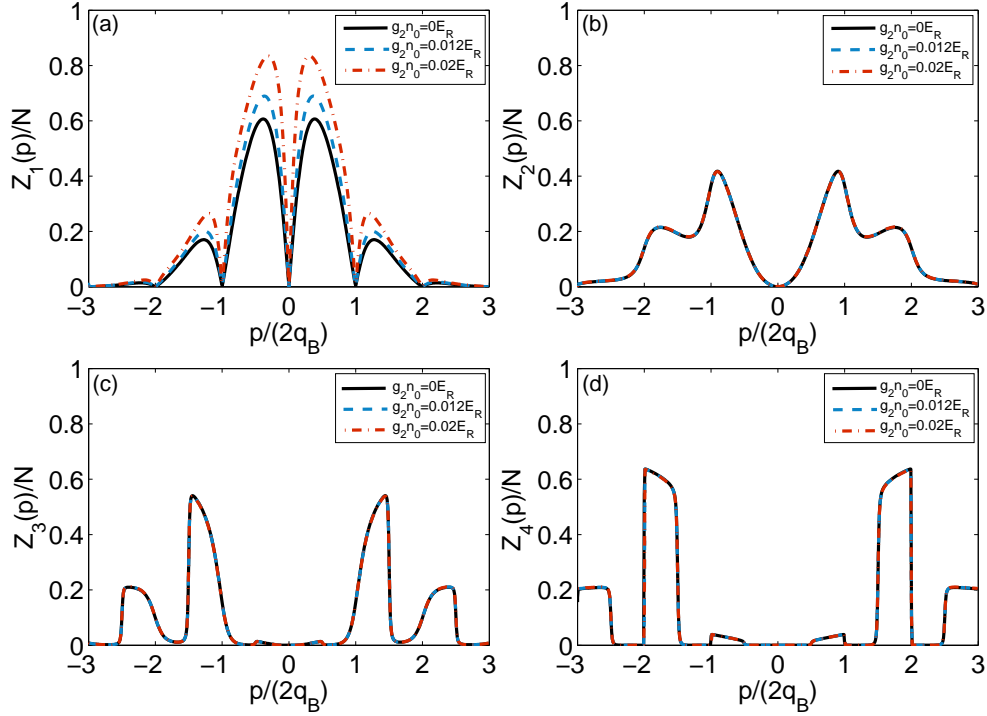}
\par\end{centering}
\caption{(Color online) The excitation strengths $Z_{j}$ $%
\left(j=1,2,3,4\right)$, as of function of momenta of an extra probe. The
full, the dashed and the dot-dashed lines corresponds to $g_{2}n_{0}=0E_{R}$%
, $g_{2}n_{0}=0.012E_{R}$ and $g_{2}n_{0}=0.02E_{R}$ with $V=5E_{R}$ and $%
g_{1}n_{0}=0.02E_{R}$, respectively.}
\label{DSF5}
\end{figure}

\begin{figure}[tbp]
\begin{centering}
\includegraphics[scale=0.7]{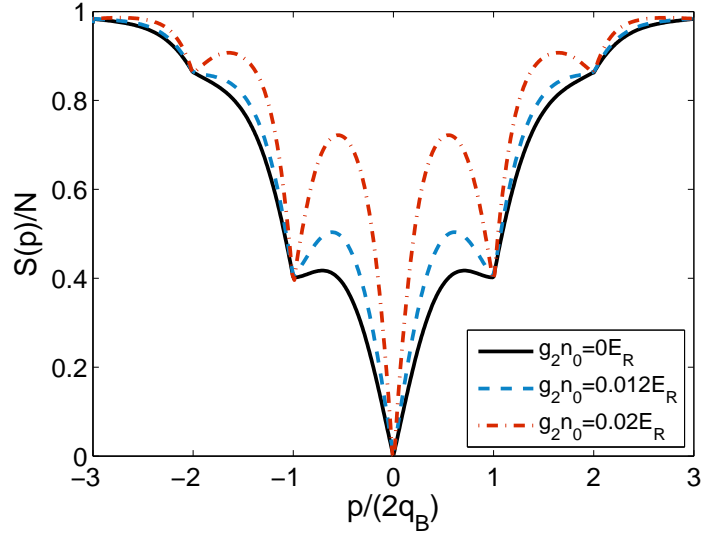}
\par\end{centering}
\caption{(Color online) Static structure factor as a function of momenta of
an extra probe. The solid, the dashed and the dot-dashed lines correspond to
$g_{2}n_{0}=0E_{R}$, $g_{2}n_{0}=0.012E_{R}$ and $g_{2}n_{0}=0.02E_{R}$ with
$V=10E_{R}$ and $g_{1}n_{0}=0.02E_{R}$, respectively.}
\label{SSF}
\end{figure}

In the first scenario where the PMI is absent ($g_{2}n_{0}=0$), both the $%
S(p,\omega )$ and $S(p)$ are plotted in bold curves in Figs. \ref{DSF10}, %
\ref{DSF5} and \ref{SSF}. In this case, we have recovered the main
conclusions of a BEC in an OL, i.e. the excitations strength $Z_{j}$ ($%
j=1,2,3,4$) toward the $j$-th Bogoliubov band develops the typical
oscillating behavior as a function of $p$ and vanishes at even multiples of
the Bragg momentum because of phononic correlations.

In the second scenario, in which the PMI is turned on ($g_{2}n_{0}\neq 0$), the
calculated lowest excitation strength $Z_{1}$ shows marked difference from
the non-PMI one, as can be clearly seen from Figs. \ref{DSF10}(a) and \ref%
{DSF5}(a). In particular, with the increase of $g_{2}n_{0}$, the developed
maximum values of $Z_{1}$ at the first Brillouin zone are greatly enhanced.
In contrast, the effects of PMI are less pronounced for higher excitation
strength $Z_{j}$ ($j=2,3,4$) (see Figs. \ref{DSF10}b, c, d and \ref{DSF5}b, c,
d). Summing up all the $Z_{j}$, we obtain the static structure factor (see
Fig. \ref{SSF}). It follows from Fig \ref{SSF} that, when the $g_{1}n_{0}$
is fixed, even a small $g_{2}n_{0}\neq 0$ will lead to an observable
modification of the static dynamic structure factor.

There are two qualitative ways to understand why the $g_{2}n_{0}$ has an
important role in determining the maximum value of the $Z_{1}$ at the first
Brillouin zone. First, this strong dependence can be explained in terms of
the effective interaction $H_{int}=\int_{-\pi }^{\pi }(c_{1}+c_{2}\cos
(x))|\psi |^{4}dx$ seen by each atom. The effective interaction is reduced
with the increasing $g_{2}n_{0}$ as emphasized in Ref. \cite{24}. Consequently, the
reduced effective interatomic interaction makes the condensate more
compressible, leading to an increase of maximum value of the $Z_{1}$ at the
first Brillouin zone. Second, following the analysis in Ref. \cite{33}, the maximum value of $Z_{1}$ close to the edge of the first
Brillouin zone can be approximately as $Z_{1}(q_{B})\sim \sqrt{\kappa
\delta /(\kappa \delta +1)}$ with $\delta =2mE_{R}/\pi ^{2}m^{\ast }$. This
simple expression shows that $Z_{1}$ is quenched both by decreasing
compressibility ($\kappa \rightarrow 0$) and by increasing the effective
mass ($\delta \rightarrow 0$). Whether such a physical picture can be
applied to our case is checked as follows. We choose to fix both $g_{1}n_{0}$
and $g_{2}n_{0}$ and plot the solid curves in Figs. \ref{DSF10}a and \ref%
{DSF5}a, corresponding to $V=10E_{R}$ and $V=5E_{R}$ respectively. The maximum of
$Z_{1}$ is increased as expected by reducing the lattice depth $V$. 
Furthermore, we can obtain the analytical expressions of the $\kappa $ and $%
m^{\ast }$ as follows ( see Eqs. (\ref{eq:9}) and (\ref{eq:10}) in
Appendix),
\begin{equation}
\frac{1}{\kappa }=c_{1}-\frac{2c_{2}^{2}}{1+4c_{1}}-\frac{2\left(
c_{2}^{2}-v^{2}\right) }{\left( 1+4c_{1}\right) ^{2}}-\frac{2\left(
c_{2}+v\right) ^{2}}{\left( 1+4c_{1}\right) ^{3}}  \label{compw}
\end{equation}%
and
\begin{equation}
\frac{1}{m^{\ast }}=1-\frac{8\left( c_{2}+v\right) ^{2}}{\left(
1+4c_{1}\right) ^{2}}  \label{massw}
\end{equation}%
It's clear from Eqs. (\ref{compw}) and (\ref{massw}) that $Z_{1}$ is
increased by the competition of both increasing $\kappa $ and $m^{\ast }$
when $g_{2}n_{0}$ increases.

Moreover, the behavior of $S(p)$ at small momenta in Fig. \ref{SSF} can be
described exactly using the sum rule approach in Eq. (\ref{lim}). As shown
in Ref. \cite{33}, the low $p$ behavior of the $S(p)$ can be
described by
\begin{equation}
\lim_{p\rightarrow 0}S(p)\sim \frac{|p|}{2\sqrt{c_{1}}}\left( 1+\frac{%
c_{2}^{2}}{c_{1}\left( 1+4c_{1}\right) }+\frac{\left( c_{2}^{2}-v^{2}\right)
}{c_{1}\left( 1+4c_{1}\right) ^{2}}-\frac{4\left( c_{2}+v\right) }{\left(
1+4c_{1}\right) ^{2}}+\frac{\left( c_{2}+v\right) }{c_{1}\left(
1+4c_{1}\right) ^{3}}\right) .
\end{equation}%
From this, we again conclude that the increase of $g_{2}n_{0}$ enhances the
value of the static structure factor at low values of $p$, as clearly shown
in Fig. \ref{SSF}.

\section{Experimental Perspective and Conclusion}

\label{sec:Summary} Summarizing, we have studied the effects of PMI on both
the sound speed and the dynamical structure factor of a BEC with PMI in an
OL. Our results of sound speed show that the PMI can strongly influence the
sound speed of BEC. Such effects of PMI can be probed experimentally by
using the Bragg spectroscopy, which directly measures the dynamic structure
factor of the system.

\section*{Acknowledgments}

We thank Ying Hu and Biao Wu for helpful and motivating discussions. This
work is supported by the NSF of China (grant nos. 11004200 and 11274315).

\appendix

\section{Sound velocity of a BEC with PMI in an OL: perturbation approach}

\subsection{Weak-potential regime}

In the weak potential and interaction regime where $v\thicksim
c_{2}\thicksim \lambda $ ($\lambda $ is a small parameter), both the OL
potential ($v$) and PMI ($c_{2}$) can be treated as a perturbation to an
unperturbed system consisting of a homogeneous qusi-1D BEC. For the
considered case in our work where PMI has the same period with OL, we could
develop a perturbation theory through the expansion of the condensate wave
function $\psi \left( x\right) $ up to second order of the small parameters,
i.e.
\begin{equation}
\psi \left( x\right) =\psi ^{\left( 0\right) }\left( x\right) +\lambda \psi
^{\left( 1\right) }\left( x\right) +\lambda ^{2}\psi ^{\left( 2\right)
}\left( x\right) +o(\lambda ^{3})  \label{eq:4}
\end{equation}%
Following the standard procedure \cite{36}, we calculate the
condensate wave function and the chemical potential order by order, and the
results are
\begin{eqnarray}
\psi \left( x\right)  &=&\sqrt{n_{0}}  \notag \\
&&-\frac{\sqrt{n_{0}}\left( V+g_{2}n_{0}\right) }{1-4k^{2}+4g_{1}n_{0}}%
\left( \left( 1-2k\right) e^{ix}+\left( 1+2k\right) e^{-ix}\right)   \notag
\\
&&+\left( A\frac{2-2k+g_{1}n_{0}}{4-4k^{2}+4g_{1}n_{0}}-B\frac{g_{1}n_{0}}{%
4-4k^{2}+4g_{1}n_{0}}\right) e^{i2x}  \notag \\
&&+\left( -A\frac{g_{1}n_{0}}{4-4k^{2}+4g_{1}n_{0}}+B\frac{2+2k+g_{1}n_{0}}{%
4-4k^{2}+4g_{1}n_{0}}\right) e^{-i2x}  \label{eq:5}
\end{eqnarray}

and%
\begin{equation}
\mu =\frac{1}{2}k^{2}+g_{1}n_{0}-\frac{\left( V+g_{2}n_{0}\right) \left(
V+3g_{2}n_{0}\right) }{1-4k^{2}+4g_{1}n_{0}}+2g_{1}n_{0}\frac{\left(
3+4k^{2}\right) \left( V+g_{2}n_{0}\right) ^{2}}{\left(
1-4k^{2}+4g_{1}n_{0}\right) ^{2}}  \label{eq:6}
\end{equation}%
with
\begin{eqnarray}
A &=&\frac{n_{0}^{1/2}V}{2}\frac{\left( 1-2k\right) \left(
V+g_{2}n_{0}\right) }{1-4k^{2}+4g_{1}n_{0}}-\frac{n_{0}^{3/2}g_{1}\left(
1-2k\right) \left( 3+2k\right) \left( V+g_{2}n_{0}\right) ^{2}}{\left(
1-4k^{2}+4g_{1}n_{0}\right) ^{2}}+\frac{n_{0}^{3/2}g_{2}}{2}\frac{\left(
3-2k\right) \left( V+g_{2}n_{0}\right) }{1-4k^{2}+4g_{1}n_{0}}  \label{eq:7a}
\\
B &=&\frac{n_{0}^{1/2}V}{2}\frac{\left( 1+2k\right) \left(
V+g_{2}n_{0}\right) }{1-4k^{2}+4g_{1}n_{0}}-\frac{n_{0}^{3/2}g_{1}\left(
1+2k\right) \left( 3-2k\right) \left( V+g_{2}n_{0}\right) ^{2}}{\left(
1-4k^{2}+4g_{1}n_{0}\right) ^{2}}+\frac{n_{0}^{3/2}g_{2}}{2}\frac{\left(
3+2k\right) \left( V+g_{2}n_{0}\right) }{1-4k^{2}+4g_{1}n_{0}}  \label{eq:7b}
\end{eqnarray}%
Furthermore, we proceed to derive the energy of the BEC, and then calculate
the effective mass $m^{\ast }$ and the compressibility $\kappa $,
respectively. The results are as follow,
\begin{equation}
\frac{1}{\kappa }=c_{1}-\frac{2c_{2}^{2}}{1+4c_{1}}-\frac{2\left(
c_{2}^{2}-v^{2}\right) }{\left( 1+4c_{1}\right) ^{2}}-\frac{2\left(
c_{2}+v\right) ^{2}}{\left( 1+4c_{1}\right) ^{3}}+o\left( 3\right)
\label{eq:9}
\end{equation}%
and
\begin{equation}
\frac{1}{m^{\ast }}=1-\frac{8\left( c_{2}+v\right) ^{2}}{\left(
1+4c_{1}\right) ^{2}}+o\left( 3\right)   \label{eq:10}
\end{equation}%
It thus follows from Eqs. (\ref{eq:9}) and (\ref{eq:10}) that the sound
speed is derived as
\begin{equation}
c_{s}=\sqrt{c_{1}}\left[ 1-\frac{c_{2}^{2}}{c_{1}\left( 1+4c_{1}\right) }-%
\frac{c_{2}^{2}-v^{2}}{c_{1}\left( 1+4c_{1}\right) ^{2}}-\frac{\left(
c_{2}+v\right) ^{2}}{c_{1}\left( 1+4c_{1}\right) ^{3}}-\frac{4\left(
c_{2}+v\right) ^{2}}{\left( 1+4c_{1}\right) ^{2}}\right]   \label{eq:11}
\end{equation}

\subsection{Tight-binding regime}

We now turn to the tight-binding regime where $v\gg c_{2}$ (while the system
is still kept in the superfluid regime). We can write the condensate wave
function $\psi \left( x\right) $ as
\begin{equation}
\psi \left( x\right) =e^{ikx}\sum_{L}e^{ikL}f\left( x-L\right)
\label{eq:12}
\end{equation}

where $L$ denotes the position of different unit cells and $f$ denotes the
Wannier functions. Having in mind that $f\left( x\right) $ is well
localized, we only take into account the overlap between the Wannier
functions associated with the nearest-neighboring sites. By substituting
Eq. (\ref{eq:12}) into Eq. (\ref{Ek}), after some straightforward algebra,
we arrived at the energy per particle reading
\begin{equation}
\varepsilon \left( k\right) =\varepsilon _{0}-\tau \cos \left( kd\right)
\label{eq:13}
\end{equation}%
Here, $\varepsilon _{0}$ is the on-site energy and $\tau $ is the tunneling
parameter. More specifically, we have calculated $\varepsilon _{0}=\frac{1}{%
2k_{L}}\int f\left( x\right) \left\{ -\frac{1}{2}\frac{\partial ^{2}}{%
\partial x^{2}}+v\cos \left( x\right) +\frac{1}{2}\left[ c_{1}+c_{2}\cos
\left( x\right) \right] df\left( x\right) ^{2}\right\} f\left( x\right) dx$.
Note that $\varepsilon _{0}$ depends on the parameters $v$, $c_{1}$ and $%
c_{2}$, but not on the wave number $k$. The tunneling parameter $\tau $ is
given by
\begin{equation}
\tau =\frac{1}{2k_{L}}\int f\left( x\right) \left\{ -\frac{1}{2}\frac{%
\partial ^{2}}{\partial x^{2}}+v\cos \left( x\right) +\frac{1}{2}\left[
c_{1}+c_{2}\cos \left( x\right) \right] df\left( x\right) ^{2}\right\}
f\left( x-d\right) dx  \label{eq:14}
\end{equation}

We must point out that during the derivation of Eq. (\ref{eq:13}), the term $%
\int f^{2}\left( x\right) f^{2}\left( x-d\right) dx$ was omitted, which for
localized functions, turn out to be much smaller, whereas the term $\int
f^{3}\left( x\right) f\left( x-d\right) dx$ was kept. Deep in tight-binding
regime, the Wannier function $f\left( x\right) $ can be approximated by a
Gaussian form $f\left( x\right) =\exp \left[ -\left( x+\frac{d}{2}\right)
^{2}/\left( 2\sigma ^{2}\right) \right] /\left( \pi ^{1/4}\sqrt{\sigma }%
\right) $, where $\sigma $ is the extension of the Gaussian which can be
determined by minimizing the energy of the system. After straightforward
calculations, the width $\sigma $ and the inverse compressibility are derived as
\begin{eqnarray}
\sigma  &\approx &\frac{d}{2\pi V^{\frac{1}{4}}}\left( 1+\frac{1}{16V^{\frac{%
1}{2}}}\right)   \label{eq:15a} \\
\frac{1}{\kappa } &\approx &\frac{1}{\sqrt{2\pi }}\left( \frac{d}{\sigma }%
\right) \left( c_{1}-c_{2}e^{-\frac{\pi ^{2}}{2}\left( \frac{\sigma }{d}%
\right) ^{2}}\right)   \label{eq:15b}
\end{eqnarray}

The effective mass can be obtained from a standard procedure \cite{36},
and here we present the final result
\begin{equation}
\frac{m}{m^{\ast }}=\left[ \frac{1}{4}\left( \frac{d}{\sigma }\right) ^{4}-%
\frac{1}{2}\left( \frac{d}{\sigma }\right) ^{2}-8\pi ^{2}Ve^{-\pi ^{2}\left(
\frac{\sigma }{d}\right) ^{2}}-8\pi \sqrt{2\pi }c_{1}\left( \frac{d}{\sigma }%
\right) e^{-\frac{1}{8}\left( \frac{d}{\sigma }\right) ^{2}}\right] e^{-%
\frac{1}{4}\left( \frac{d}{\sigma }\right) ^{2}}  \label{eq:17}
\end{equation}%
With both the compressibility $\kappa $ and the effective mass $m^{\ast }$
being derived, we can readily calculate the sound speed in the tight-binding
regime. Our analysis result is consistent with the numerical calculations
(Fig. \ref{Sound}) in corresponding regimes, suggesting that the
tight-binding treatment Eq. (\ref{eq:12}) is a reliable method.

\end{document}